\documentclass[aps,preprint,tightenlines,byrevtex,nofootinbib]{revtex4}
\usepackage{amsmath,amssymb}
\usepackage{graphicx}
\usepackage{color}
\usepackage{comment}
\usepackage{bbold}
\def\nn{\nonumber\\}

\def\eqref#1{{(\ref{#1})}}
\def\mc#1{\mathcal{#1}}
\def\Tr{\textrm{Tr}}

\def\D{\Delta}
\def\L{\Lambda}
\def\Lqcd{\Lambda_{\textrm{QCD}}}
\def\S{\Sigma}
\def\X{\Xi}
\def\O{\Omega}

\def\d{\delta}
\def\e{\epsilon}

\def\s{\sigma}

\def\ltap{\ \raise.3ex\hbox{$<$\kern-.75em\lower1ex\hbox{$\sim$}}\ }
\def\slash#1{\rlap{$#1$}/} 
\def\mreln{ \mathsf{M} }
\def\nlo{\rm NLO}
\def\nnlo{\rm NNLO}
\def\aone{a_{1}}
\def\atwo{a_{2}}
\def\bone{b_{1}}
\def\btwo{b_{2}}
\def\GMO{\rm GMO}

\begin{document}

{\count255=\time\divide\count255 by 60 \xdef\hourmin{\number\count255}
  \multiply\count255 by-60\advance\count255 by\time
  \xdef\hourmin{\hourmin:\ifnum\count255<10 0\fi\the\count255}}

\preprint{NT-LBNL-11-014}
\preprint{UCB-NPAT-11-010}

\title{Evidence for non-analytic light quark mass dependence in the baryon spectrum}

\author{Andr\'{e} Walker-Loud}
\email[]{awalker-loud@lbl.gov}
\affiliation{Nuclear Science Division, Lawrence Berkeley National Laboratory, Berkeley, CA 94720 \vspace{4pt} }

\date{\today}

\vskip1.5in
\begin{abstract}
Using precise lattice QCD computations of the baryon spectrum, we present the first direct evidence for the presence of contributions to the baryon masses which are non-analytic in the light quark masses; contributions which are often denoted \textit{chiral logarithms}.  We isolate the poor convergence of $SU(3)$ baryon chiral perturbation theory to the flavor-singlet mass combination.  The flavor-octet baryon mass splittings, which are corrected by chiral logarithms at next to leading order in $SU(3)$ chiral perturbation theory, yield baryon-pion axial coupling constants $D, F, C$ and $H$ consistent with QCD values; the first evidence of chiral logarithms in the baryon spectrum.  The Gell-Mann--Okubo relation, a flavor-\textbf{27} baryon mass splitting, which is dominated by chiral corrections from light quark masses, provides further evidence for the presence of non-analytic light quark mass dependence in the baryon spectrum; we simultaneously find the GMO relation to be inconsistent with the first few terms in a taylor expansion in $m_s - m_l$, which must be valid for small values of this $SU(3)$ breaking parameter.  Additional, more definitive tests of $SU(3)$ chiral perturbation theory will become possible with future, more precise, lattice calculations.
\end{abstract}

\maketitle

%
\section{Introduction}
%
Quantum chromodynamics (QCD) is one of the fundamental gauge theories of the standard model of particle physics, encoding the interactions amongst quarks and gluons.
At high energies, the theory exhibits the property of asymptotic freedom where the coupling between the quarks and gluons runs to zero as the interactions are probed with larger momentum transfer.
Conversely, at low energies, at a scale of $\Lqcd \sim 1$~GeV, the coupling between the quarks and gluons becomes $\mc{O}(1)$, and the theory is no longer amenable to a perturbative treatment; the quark and gluon degrees of freedom are bound into the observed hadronic degrees of freedom, the protons, neutrons, pions, etc., which leave only subtle clues about the underlying fundamental theory of QCD.

These properties of QCD, as well as many others, are now well established thanks to a variety of techniques that have been developed to understand the rich phenomena that emerge from the theory.
One of the most important tools is lattice QCD, a numerical solution to the theory, performed on a discrete, Euclidean space-time lattice.
With algorithmic advances and ever growing computing power, state of the art lattice QCD calculations are performed at several lattice spacings, with moderate physical space-time volumes and with dynamical light quark masses at or near their physical values~\cite{Aoki:2008sm,Bazavov:2009bb,Aoki:2009ix,Durr:2010vn,Durr:2010aw}.
Recently, the ground state hadron spectrum, composed of up, down and strange quarks, has been reproduced from lattice calculations with a few percent uncertainty~\cite{Durr:2008zz}.  This serves as an important benchmark in demonstrating the ability for these numerical calculations to produce precise quantitative predictions for hadronic physics observables.
Indeed, lattice calculations are playing an important role in many areas of both nuclear and high energy physics~\cite{LattProc}.

In addition to this numerical solution to QCD, a variety of analytic methods have been developed to understand the low-energy regime of the theory.  The most prominent method is chiral perturbation theory ($\chi$PT) which exploits an approximate global symmetry of QCD~\cite{Weinberg:1978kz}.
For the $up$, $down$ and $strange$ quarks with masses less than $\Lqcd$, the QCD Lagrangian is approximately invariant under global chiral transformations of the quark fields such that the theory has an approximate $SU(3)_L \times SU(3)_R$ chiral symmetry, which becomes exact in the limit the quarks are massless.
This approximate chiral symmetry is spontaneously broken to the $SU(3)_V$ subgroup by the QCD vacuum giving rise to the pion octet pseudoscalar pseudo--Nambu-Goldstone bosons, the pions, kaons and eta.
The realization of this chiral symmetry, as well as its spontaneous and explicit symmetry breaking, can be described by constructing a chiral Lagrangian which contains this pion octet as well as other $SU(3)_V$ hadron multiplets as explicit degrees of freedom.
In the chiral limit, the pion octet become exact Nambu-Goldstone bosons which have only derivative couplings to themselves and other hadrons.
This theory, $\chi$PT, is non-renormalizable and contains an infinite number of operators whose forms are constrained by the global symmetries of QCD, while the coefficients of these operators, the low energy constants (LECs), are unconstrained and must be determined by comparing with experimental data and/or the results of numerical lattice QCD calculations.
The quantitative relevance of these operators are dictated by an expansion in the soft momentum of the pion octet and the light quark masses suppressed by the chiral symmetry breaking scale, $\L_\chi$;
hadronic observables can be computed at low energies to any fixed precision by keeping operators to a given order in the chiral expansion, thus requiring determination of only a finite number of the LECs.

One of the principle applications of $\chi$PT has been to determine the light quark mass dependence of various hadronic observables, the simplest of which is the light hadron spectrum.
The motivation comes from the significant numerical cost of performing lattice QCD calculations at the physical values of the $up$ and $down$ quark masses.  $\chi$PT can be used to extrapolate the numerical lattice QCD results to the physical values of the light quark masses, in the process determining some of the LECs associated with the quark mass dependent operators.  This program has been very successful when applied to the Nambu-Goldstone meson spectrum and decay constants, see Ref.~\cite{Colangelo:2010et} for a review, beginning with the first significant comparison of lattice QCD results with $\chi$PT~\cite{Aubin:2004fs}.

The comparison with the light baryon spectrum has been wrought with more significant difficulties and the overall convergence, and usefulness of the $SU(3)$ baryon $\chi$PT is in question.
These challenges are not unexpected; first, there is a dense spectrum of low lying excited states, introducing new scales in the theory; second, while the expansion parameters of $\chi$PT in the Nambu-Goldstone meson sector are given by $\e_m \sim m_{K,\pi,\eta}^2 / \L_\chi^2$, the small expansion parameter when the baryon fields are included becomes $\e \sim m_{K,\pi,\eta} / \L_\chi$~\cite{Gasser:1987rb}.  For the physical kaon, $\e \simeq 1/2$ and from general expectations of asymptotic series, one does not expect this theory to have a controlled perturbative expansion.
A few recent comparisons of $SU(3)$ baryon $\chi$PT to numerical lattice QCD results have led to the conclusion the three flavor chiral expansion is failing to provide a controlled, convergent expansion~\cite{WalkerLoud:2008bp,Ishikawa:2009vc,Torok:2009dg}.
The issues of convergence are not limited to the three flavor expansion~\cite{Beane:2004ks} and recent analysis indicates the range of expansion of the two flavor theory, considering only an expansion about the limit of vanishing up and down quark masses, extends only to $m_\pi^{max} \simeq 300$~MeV~\cite{WalkerLoud:2008bp,WalkerLoud:2008pj}.

These challenges have led to a number of efforts to reorganize the expansion for baryon $\chi$PT.  The initial approach is known as heavy baryon $\chi$PT (HB$\chi$PT) which treats the baryons as nearly static fields allowing for an expansion in inverse powers of the baryon mass~\cite{Jenkins:1990jv,Jenkins:1991es}, modeled after the heavy quark effective theory~\cite{Georgi:1990um}.  This led to significant phenomenological successes which are partly reviewed in Refs.~\cite{Bernard:1995dp,Bernard:2007zu}.
Early on, it was recognized the convergence of the theory would be problematic because of the large contributions from kaon and eta loops in various observables.  A new regularization scheme was proposed, the introduction of a (chiral symmetry violating) long range regulator, eg. a dipole regulator, to soften the contribution from the kaon and eta loops~\cite{Donoghue:1998bs}.
When applied to chiral extrapolations of lattice QCD results, this led to some successes in simultaneously describing both the numerical results and physical observables~\cite{Leinweber:1999ig,Leinweber:2003dg}.
An additional reorganization of the chiral expansion, equivalent to a resummation of the leading kinetic corrections to the baryon propagators was constructed and has become known as infrared regularized baryon $\chi$PT~\cite{Becher:1999he}; with several offshoots to deal with renormalization of higher loop corrections~\cite{Fuchs:2003qc,MartinCamalich:2010fp}.
Lattice QCD calculations in the last few years have also made feasible the use of the $SU(2)$ expansion for hyperons~\cite{Tiburzi:2008bk,Jiang:2009fa}.
In this work, we further examine a new application of an old idea: combining the large $N_c$ expansion~\cite{'tHooft:1973jz,Witten:1979kh} with the $SU(3)$ chiral expansion~\cite{Dashen:1993as,Dashen:1993ac,Jenkins:1993zu,Dashen:1993jt,FloresMendieta:2000mz}.
This approach has a few formal advantages over the other methods.  In the large $N_c$ limit, there is an extra symmetry, the contracted spin-flavor symmetry~\cite{Dashen:1993as,Dashen:1993ac}, allowing for an unambiguous field-theoretic method to include the low lying decuplet baryon resonances in the theory; in the large $N_c$ limit, the spin-$1/2$ and -$3/2$ baryons become degenerate and infinitely heavy.
Further, while the large $N_c$ and $SU(3)$ chiral expansions on their own may not provide well converged effective theories, the combined expansions may prove sufficient for a controlled perturbative expansion.
This approach was first explored in Ref.~\cite{Jenkins:2009wv} where it was demonstrated the predictions from the combined large $N_c$ and $SU(3)$ expansions on the baryon spectrum are well met for a range of light quark masses.

Having a controlled expansion is necessary but not sufficient to claim success.  The principle prediction from $\chi$PT are the contributions to hadronic observables which are non-analytic in the light quark masses.
The masses of Nambu-Goldstone boson is given to leading order by the Gell-Mann-Oakes--Renner Relation~\cite{GellMann:1968rz}, $m_{i,j}^2 = B(m_i + m_j)$, with a meson composed of a quark--anti-quark pair of (anti) flavors $i$ and $j$ and $m_i$ is the mass of a quark with flavor $i$.  Therefore, in $\chi$PT, the non-analytic light quark mass dependence arises from pion-octet loops, which often contribute $\ln(m_{K,\pi,\eta}^2)$ terms to hadronic observables, and are commonly referred to as \textit{chiral logs}. 
These contributions can not arise from a finite number of local counterterms but only from the long range contributions from the light pion octet degrees of freedom, the \textit{pion cloud}. 
Isolating this predicted light quark mass dependence in lattice QCD results has been a major challenge for many years.  The definitive identification of these contributions is hailed as a signal that the $up$ and $down$ (and $strange$) quarks are sufficiently light that the lattice results can be described accurately by $\chi$PT.
This task has proved to be very challenging, as often, these non-analytic light quark mass contributions are subleading, or masked by other systematics.

\bigskip
In this work, we present for the first time, direct evidence of non-analytic light quark mass dependence in the baryon spectrum.
As will be discussed in this article, this work is not the definitive work on the subject, as there are many systematics which must be resolved, but this is an important first step in the quest for chiral logs.

%
\section{The heavy baryon chiral Lagrangian and the large $N_c$ expansion \label{sec:largeNc}}
%

%
\subsection{Heavy Baryon Chiral Lagrangian in the $1/N_c$ Expansion}
%
The three flavor heavy baryon chiral Lagrangian at leading order (LO) in the momentum expansion and to first order in the chiral-symmetry breaking quark mass matrix ${\cal M}_q \equiv {\rm diag}(m_u, m_d, m_s)$ is given by~\cite{Jenkins:1990jv,Jenkins:1991es},
\begin{align}\label{eq:HBLag}
\mathcal{L} =&\ \Tr\, \bar{B}_v \left( i v \cdot {\cal D} \right) B_v
	-\bar{T}_v^\mu \left( i v \cdot {\cal D}\right) T_{v \, \mu} - \frac{1}{4} \D_0 \ Tr \ \bar{B}_v B_v 
	+ \frac{5}{4} \D_0 \ \bar{T}_v^\mu T_{v \, \mu}
\nonumber\\&
	+2D\, \Tr \left( \bar{B}_v S_v^\mu \left\{\mc{A}_\mu, B_v \right\} \right)
	+2F\, \Tr \left( \bar{B}_v S_v^\mu \left[ \mc{A}_\mu, B_v \right] \right)
\nonumber\\&
	+\mc{C}\, \left( \bar{T}_v^\mu \mc{A}_\mu B_v + \bar{B}_v \mc{A}_\mu T_v^\mu \right)\, 
         +2 \mc{H}\, \bar{T}_v^\mu S_v^\nu \mc{A}_\nu T_{v \, \mu}
\nonumber\\&	
	+2\s_B \ \Tr \left( \bar{B}_v B_v \right) \Tr  \mc{M}_+ 
	-2\s_T\ \bar{T}_v^\mu T_{v \, \mu} \Tr  \mc{M}_+ 
\nonumber\\&
	+2b_D \Tr \left( \bar{B}_v \left\{\mc{M}_+,  B_v\right\} \right)
	+2b_F \Tr \left( \bar{B}_v \left[\mc{M}_+,  B_v \right]  \right)
	+2b_T\, \bar{T}_v^\mu \mc{M}_+ T_{v \, \mu}
\end{align}
where the spin-$1/2$ octet baryon fields $B_v$ and spin-$3/2$ decuplet baryon fields $T_v^\mu$ are two-component velocity-dependent baryon fields which are related to the usual four-component relativistic Dirac spin baryon fields $B$ and $T^\mu$ by 
\begin{align}
	B_v (x) &= \frac{1 + \slash v}{2} e^{i M_0 v \cdot x} B(x), \nonumber\\
	T_v^\mu (x) &= \frac{1 + \slash v}{2} e^{i M_0 v \cdot x} T^\mu(x)\, .
\end{align}
The mass $M_0$ is the flavor-singlet mass of the baryon octet and decuplet baryons in the $SU(3)$ chiral limit $m_q \rightarrow 0$.  Specifically, 
\begin{equation}
M_0 = \frac{5}{4} \langle M_{\bf 8} \rangle - \frac{1}{4} \langle M_{\bf 10} \rangle ,
\end{equation}
where $\langle M_{\bf 8} \rangle$ and $\langle M_{\bf 10} \rangle$ are the average flavor-singlet masses of the spin-$1 / 2$ flavor-octet baryons and the spin-$3 / 2$ flavor-decuplet baryons, respectively, in the chiral limit.
In the large $N_c$ expansion, $M_0$ is $\mc{O}(N_c)$ for baryons with $N_c$ quarks.  
The leading heavy baryon chiral Lagrangian also contains the flavor-singlet hyperfine mass splitting
\begin{equation}
\Delta_0 = \langle M_{\bf 10} \rangle -  \langle M_{\bf 8} \rangle,
\end{equation}  
which is proportional to the total spin-squared $J_v^2$ of each baryon multiplet.  The mass parameter $\Delta_0$ is $\mc{O}(1/N_c)$ in the $1/N_c$ expansion. 
The $SU(3)$ flavor representations of the QCD baryons are
the flavor-octet
\begin{equation}
B = \begin{pmatrix}
	\frac{1}{\sqrt{2}}\S^0 + \frac{1}{\sqrt{6}} \L & \S^+& p \\
	\S^-& -\frac{1}{\sqrt{2}}\S^0 + \frac{1}{\sqrt{6}} \L & n \\
	\Xi^-& \Xi^0 & -\frac{2}{\sqrt{6}} \L
\end{pmatrix},
\end{equation}
and the completely symmetric rank-3 flavor-decuplet $T_{ijk}$, normalized such that $T_{uuu} = \Delta^{++}$.
The heavy baryon chiral Lagrangian also
contains four independent baryon-pion couplings, the axial couplings $D$, $F$, $\mc{C}$ and $\mc{H}$.  The couplings $D$ and $F$ describe the usual baryon-octet pion couplings; $\mc{C}$ describes pion couplings between octet and decuplet baryons; and $\mc{H}$ describes the pion coupling of the decuplet baryons.  The pion octet fields 
\begin{align}
\Pi \equiv \pi^a T^a = \begin{pmatrix}
	\frac{1}{\sqrt{2}}\pi^0 + \frac{1}{\sqrt{6}} \eta & \pi^+& K^+ \\
		\pi^-& -\frac{1}{\sqrt{2}}\pi^0 + \frac{1}{\sqrt{6}} \eta & K^0 \\
		K^-& \bar{K^0}& -\frac{2}{\sqrt{6}} \eta
	\end{pmatrix}
\end{align}
appear in the heavy baryon chiral Lagrangian in the nonlinear representation $\xi^2 = \S = e^{2i\Pi/f}$, where $f \sim 130$~MeV is
the pion decay constant in the chiral limit.  The vector and axial vector pion combinations
\begin{eqnarray}
\mc{A}_\mu &=& \frac{i}{2} \left( \xi \partial_\mu \xi^\dagger - \xi^\dagger \partial_\mu \xi \right)\, , \nonumber\\
\mc{V}_\mu &=& \frac{1}{2}  \left( \xi \partial_\mu \xi^\dagger + \xi^\dagger \partial_\mu \xi \right),
\end{eqnarray}
appear in the baryon-pion couplings and through the  
baryon covariant derivative $\mc{D}_\mu = \partial_\mu + i \mc{V}_\mu$.  
In the heavy baryon chiral Lagrangian, $S_v^\mu$ is the spin operator which acts on the spinor portion of the baryon field.

Additional dependence on the pion field enters through the quark mass matrix spurion   
\begin{equation}
\mc{M}_+ = \frac{1}{2} \left( \xi {{\cal M}_q}^\dagger \xi + \xi^\dagger {\cal M}_q \xi^\dagger \right)\, .
\end{equation}
In this work, we compare with lattice computations performed with degenerate $u$ and $d$ quark masses $m_u = m_d = m_l$, so
the quark mass matrix reduces to 
\begin{equation}\label{eq:Mq18}
{\cal M}_q = \frac{1}{3} \left(2m_l + m_s \right) \openone + \frac{2}{\sqrt{3}} \left(m_l - m_s\right) T^8\, .
\end{equation}
There are two flavor-singlet contributions to the baryon masses with one insertion of the quark mass matrix coming from the terms proportional to $\s_B$ and $\s_T$.  
There are also three flavor-octet contributions to the baryon masses with a single insertion of the quark mass matrix, proportional to $b_D$, $b_F$ and $b_T$ (called $b_C$ previously~\cite{Jenkins:1991ts}).

The $1/N_c$ expansion~\cite{'tHooft:1973jz} for baryons~\cite{Witten:1979kh} leads to the emergence of a spin-flavor symmetry~\cite{Dashen:1993as,Dashen:1993ac,Dashen:1993jt} for large-$N_c$ baryons.  
In Ref.~\cite{Jenkins:1995gc}, the heavy baryon Lagrangian was formulated in the $1/N_c$ expansion.  Relations amongst the coefficients in the heavy baryon chiral Lagrangian occur at leading and subleading orders in the $1/N_c$ expansion, which reduces the number of independent chiral coefficients in the heavy baryon chiral Lagrangian at leading and subleading orders in $1/N_c$.  
In addition, there exists a planar flavor symmetry~\cite{Jenkins:1995gc} at leading order in $1/N_c$, which relates flavor-singlet to flavor-octet parameters at this order, further reducing the number of independent chiral coefficients in the heavy baryon chiral Lagrangian at leading order in the $1/N_c$ expansion.
In particular, planar QCD flavor symmetry relates the flavor-singlet quark mass parameters $\s_B$ and $\s_T$ to the flavor-octet quark mass parameters $b_D$, $b_F$ and $b_T$ at leading orders in $1/N_c$.     
The flavor-octet and flavor-singlet quark mass parameters are given in terms of the coefficients $b_{n}$ of the spin-0 flavor-octet $1/N_c$ expansion,%
\footnote{Here, we adopt a simplified notation for the operator coefficients compared to Ref.~\cite{Jenkins:1995gc}.} 
where the subscript $n$ refers to the fact that the corresponding operator $\mc{O}_{(n)}$ is an $n$-body quark operator which is accompanied by an explicit factor of $N_c^{1-n}$.
To first subleading order in the $1/N_c$ expansion, the mass matrix parameters of the heavy baryon chiral Lagrangian for QCD with $N_c=3$ are given by
\begin{align}\label{eq:lgNnlomQ}
&b_D = \frac{1}{4} b_{2}\, ,&
&b_F = \frac{1}{2} b_{1} + \frac{1}{6} b_{2}\, ,&
&b_T = -\frac{3}{2} b_{1} -\frac{5}{4} b_{2}\, ,&
\nonumber\\
&\s_B = \frac{1}{2} b_{1} + \frac{1}{12} b_{2}\, ,&
&\s_T = \frac{1}{2} b_{1} + \frac{5}{12} b_{2}\, .&
\end{align}
The axial couplings $D$, $F$, $\mc{C}$ and $H$ also have an expansion in terms of spin-1 flavor-octet coefficients $a_{n}$ of the $1/N_c$ expansion.  To first subleading order in $1/N_c$, the pion-baryon couplings of the heavy baryon chiral Lagrangian for QCD with $N_c=3$ are related to the $1/N_c$ coefficients by~\cite{Dashen:1994qi,Jenkins:1995gc}%
\footnote{The $1/N_c$ operator analysis has recently been extended to the two-body axial current operators~\cite{Lutz:2010se}, such as 
$\Tr \left( \bar{B} \mc{A} \cdot \mc{A} B \right)$.} 
\begin{align}\label{eq:lgNnloA}
&D = \frac{1}{2}a_{1}\, ,&
&F = \frac{1}{3}a_{1} + \frac{1}{6}a_{2}\, ,&
\nonumber\\
&\mc{C} = -a_{1}\, ,&
&\mc{H} = -\frac{3}{2}a_{1} - \frac{3}{2}a_{2}\, .&
\end{align}

%
\subsection{Mass Relations $R_1$, $R_3$ and $R_4$}
%

In Ref.~\cite{Jenkins:2009wv}, it was argued a better approach to exploring the baryon spectrum was to utilize our knowledge of both large $N_c$ as well as $SU(3)$ symmetry which is known to work well for the experimental spectrum~\cite{Jenkins:1995td}; instead of considering the individual baryon masses directly, one should explore the light quark mass dependence of various linear combinations of the baryon masses, chosen to have definite scaling in terms of $1/N_c$ and $SU(3)$ symmetry breaking.%
\footnote{Ref.~\cite{Semke:2011ez} utilized the large $N_c$ relations between operators in baryon $\chi$PT to study the baryon spectrum, but not the linear combinations constructed to have definite scaling in $m_s - m_l$ and $1/N_c$.} 
The various linear combinations were determined in Ref.~\cite{Jenkins:1995td}.
In Ref.~\cite{Jenkins:2009wv}, it was demonstrated that the predicted scaling with both $1/N_c$ and $\left(m_s - m_l\right)$ was clearly visible in the lattice data.  The first few mass combinations had statistically meaningful values over the range of quark masses, but there were not enough statistics to resolve all of them.
In this work, we focus our attention on three of these mass relations, $R_1, R_3$ and $R_4$.%
\footnote{The relation $R_2$ gives at leading order the hyper-fine splitting $\Delta_0$.  For the current lattice data set, this quantity provides no further information over the use of $R_1$.}
These mass relations are given by
\begin{equation}
R_i = \frac{\sum_j c_{ij} M_j}{\sum_j |c_{ij}|}
\end{equation}
where
\begin{align}
&\mreln_1 = \sum_j c_{1j} M_j = 25(2M_N +M_\L + 3M_\S +2M_\X) - 4(4M_\D +3M_{\S^*} +2M_{\X^*} +M_\O)\, ,&
\nn
&\mreln_3 = \sum_j c_{3j} M_j = 5(6M_N +M_\L - 3M_\S - 4M_\X) - 2(2M_\D - M_{\X^*} - M_\O)\, ,&
\nn
&\mreln_4 = \sum_j c_{4j} M_j = M_N + M_\L - 3M_\S + M_\X\, ,&
\end{align}
and for example $R_4 = \mreln_4 / 6$.

These relations are designed to isolate various operators in the combined $1/N_c$ and $SU(3)$-breaking expansions.  
At $\mc{O}(m_q)$, only relations $R_1$--$R_4$ are non-vanishing.  
For this reason, the relations $R_5$ -- $R_8$ are particularly interesting to use with light quark mass extrapolations, as the leading contribution begins with the chiral loops at $\mc{O}(m_q^{3/2})$.  However, even more precise results of the baryon spectrum than exist are needed for these relations.
Using the large $N_c$ expansions through second non-trivial order, and working through next-to-leading order (NLO) in the chiral expansion, the relation $R_1$ is given by
\begin{align}\label{eq:R1_nlo}
\frac{3}{2}\ R_1(m_l,m_s) =&\ M_0
	-\left( \frac{3}{4}\bone + \frac{5}{24}\btwo \right) (2m_l + m_s)
\nonumber\\&
	-\frac{1}{12}{\left( 35 \aone^2 -5 \atwo^2 \right)}  \left(
		\frac{3\mc{F}(m_\pi,0,\mu)+4 \mc{F}(m_K,0,\mu)+\mc{F}(m_\eta,0,\mu)}{8(4\pi f)^2} \right)
\nonumber\\&
		-\frac{1}{12}{\aone^2}  \bigg[ 50 \left( \frac{3\mc{F}(m_\pi,\D,\mu)
		+4\mc{F}(m_K,\D,\mu)+\mc{F}(m_\eta,\D,\mu)}{8(4\pi f)^2} \right) 
\nonumber\\&\qquad
		-4 \left( \frac{3\mc{F}(m_\pi,-\D,\mu)+4\mc{F}(m_K,-\D,\mu)+\mc{F}(m_\eta,-\D,\mu)}{8(4\pi f)^2} \right)
	\bigg]
%
\end{align}
The non-analytic function is defined as
\begin{multline}\label{eq:F}
\mc{F}(m,\D,\mu) =
	(\D^2 - m^2 +i\e)^{3/2}
	\ln \left( \frac{\D+\sqrt{\D^2 - m^2 + i\e}}{\D - \sqrt{\D^2 - m^2 + i\e}} \right)
\\
	-\frac{3}{2}\D m^2 \ln \left( \frac{m^2}{\mu^2}\right)
	-\D^3 \ln \left(\frac{4\D^2}{m^2}\right)\, .
\end{multline}
which has the limits and properties
\begin{align}
\mc{F}(0,\D,\mu) &= 0
\nonumber\\
\mc{F}(m,0,\mu) &= \pi m^3
\nonumber\\
\mc{F}(m,-\D,\mu)&= \left\{ \begin{array}{lc}
	-\mc{F}(m,\D,\mu) +2i\pi (\D^2 - m^2)^{3/2}, & m < |\D| \\
	-\mc{F}(m,\D,\mu) +2\pi (m^2 - \D^2)^{3/2}, & m > |\D|
	\end{array} \right. \, .
\end{align}
For the baryon spectrum, the leading non-analytic light quark mass dependence is encoded in this function.
As such, it is of particular interest to find evidence of this behavior in the spectrum.

The mass relations $R_3$ and $R_4$ vanish in both the $SU(3)$ chiral and vector limits, making them more sensitive to the NLO non-analytic light quark mass dependence.  At NLO in the chiral expansion, and to the first two non-trivial orders in the large $N_c$ expansion, these relations are given by
\begin{align}\label{eq:R3nlo}
39 \ R_3(m_l,m_s) =&\ {20} \ \bone \left(m_s - m_l \right)
\nonumber\\&
	-\frac{1}{3}{\left( 20 \aone^2 - 5 \atwo^2 \right)}
		 \left( \frac{3\mc{F}(m_\pi,0,\mu)-2 \mc{F}(m_K,0,\mu)-\mc{F}(m_\eta,0,\mu) }{(4\pi f)^2} \right)
\nonumber\\&
		-\frac{1}{3}{\aone^2} \bigg[ 
		35 \left( \frac{3\mc{F}(m_\pi,\D,\mu)-2\mc{F}(m_K,\D,\mu)-\mc{F}(m_\eta,\D,\mu)}{(4\pi f)^2} \right)
\nonumber\\&\qquad\quad
		-\left( \frac{3\mc{F}(m_\pi,-\D,\mu)
			-2\mc{F}(m_K,-\D,\mu)
			-\mc{F}(m_\eta,-\D,\mu)}{(4\pi f)^2} \right)
	\bigg]\, ,
\end{align}
\begin{align}\label{eq:R4nlo}
R_4(m_l,m_s) =& -\frac{5}{18}	\ \btwo \left( m_s - m_l \right)
\nonumber\\&\
	+\frac{1}{36}{\left( \aone^2+ 4\aone \atwo +\atwo^2\right)} \left( 
	\frac{3\mc{F}(m_\pi,0,\mu)-2 \mc{F}(m_K,0,\mu)-\mc{F}(m_\eta,0,\mu)}{(4\pi f)^2} \right)
\nonumber\\&\
	-\frac{2}{9} {\aone^2} \left(
		\frac{3\mc{F}(m_\pi,\D,\mu)-2\mc{F}(m_K,\D,\mu)-\mc{F}(m_\eta,\D,\mu)}{(4\pi f)^2}
	\right)\, .
\end{align}

In addition to these three mass relations, we also explore the Gell-Mann--Okubo relation
\begin{equation}\label{eq:GMO}
\D_{\GMO} = \frac{3}{4}M_\L + \frac{1}{4}M_\S - \frac{1}{2}M_N - \frac{1}{2}M_\X\, .
\end{equation}
Since the quark mass operator contains pieces which transform as both an $\mathbf{8}$ as well as a $\mathbf{1}$ under $SU(3)$ transformations, Eq.~\eqref{eq:Mq18}, there are non-vanishing contributions to the GMO relation.
However, mass operators which transform as an $\mathbf{8}$ make vanishing contributions to Eq.~\eqref{eq:GMO}.  
The leading mass operator which makes a non-zero contribution to the GMO relation transforms as a flavor-$\mathbf{27}$.
These corrections can arise either from chiral loops or from a mass operator containing two or more quark mass insertions.
This makes the GMO relation particularly interesting to explore with lattice QCD calculations; the leading contribution to this mass relation comes from chiral loop effects which are non-analytic in the light quark masses.
Experimentally, the GMO relation is found to be
\begin{equation}\label{eq:GMO_experiment}
	\D_{\GMO}^{phy} = 6.45 \textrm{ MeV}\, .
\end{equation}
Each baryon mass in the relation receives non-analytic mass corrections which scale as $\d M_B \propto N_c m_s^{3/2}$.  These large corrections may lead to the expectation that the GMO relation receives large contributions from the loop corrections.  However, one can show these $N_c m_s^{3/2}$ terms are proportional to $\mathbf{1}$ under $SU(3)$ transformations.  Additionally, the $m_s^{3/2}$ contributions transform as an $\mathbf{8}$ while the $m_s^{3/2} / N_c$ corrections transform as a flavor-$\mathbf{27}$.
This provides an extra $1/N_c^2$ on top of the chiral suppression, explaining the relatively small value of the GMO relation~\cite{Jenkins:1995gc}.

At next-to-leading order in the chiral and large $N_c$ expansions, the Gell-Mann--Okubo relation is
\begin{align}\label{eq:GMOnlo}
\D_{\GMO}^{\nlo} =&\ \frac{a_{1}^2}{36(4\pi f)^2} 
	\bigg[ \mc{F}(m_\pi,0,\mu)
		-4\mc{F}(m_K,0,\mu) 
		+ 3\mc{F}(m_\eta,0,\mu)
\nonumber\\&\qquad\qquad\qquad
	 	+2 \mc{F}(m_\pi,\D,\mu) 
		- 8 \mc{F}(m_K,\D,\mu) 
		+ 6\mc{F}(m_\eta,\D,\mu)
	\bigg]
\nonumber\\&\ 
	+\frac{4 a_{1} a_{2} + a_{2}^2}{36(4\pi f)^2}
	\bigg[ \mc{F}(m_\pi,0,\mu)
		-4\mc{F}(m_K,0,\mu) 
		+ 3\mc{F}(m_\eta,0,\mu)	
	\bigg]\, ,
\end{align}

In this article, we will also be interested in the next-to-next-to-leading order (NNLO) formula.  This can be determined from Ref.~\cite{WalkerLoud:2004hf,WalkerLoud:2006sa}.  Retaining the subleading $N_c$ relations for the quark mass operators, Eq.~\eqref{eq:lgNnlomQ}, but only the leading relations for the axial couplings ($a_2 \rightarrow 0$), the NNLO contributions to the GMO formula are
\begin{align}\label{eq:GMOnnlo}
\D_{\GMO}^{\nnlo} =&\ 
	\frac{1}{48}\frac{\left( b^M_3 + b^M_4 \right)\left(m_s - m_l\right)^2}{4\pi f}
	+\frac{23 \aone^2}{384 \, M_0}
	\left(\frac{ m_\pi^4 - 4m_K^4 + 3m_\eta^4 }{(4\pi f)^2} \right)
\nonumber\\&\
	+\left( \frac{b^A_3}{96\pi f} 
		+\frac{13 \aone^2 }{192 \, M_0}
	\right)
		\left( \frac{m_\pi^4 \ln(m_\pi^2 / \mu^2) 
			-4m_K^4 \ln (m_K^2 / \mu^2)
			+3m_\eta^4 \ln(m_\eta^2 / \mu^2)}
		{(4\pi)^2 f^2} \right)
\nonumber\\&\ 
	+\frac{\aone^2}{16\pi^2 f^2} \bigg\{
		\bone \bigg[ 
			\frac{2m_l+m_s }{12} 
				\Big(  \mc{J}(m_\pi,\D,\mu) 
					- 4\mc{J}(m_K,\D,\mu) 
					+ 3\mc{J}(m_\eta,\D,\mu)
\nonumber\\&\qquad\qquad\qquad
					+m_\pi^2 -4m_K^2 +3m_\eta^2 \Big)
		\bigg]
\nonumber\\&\qquad
		+\btwo \bigg[ 
			\frac{-9m_s + 16 m_l}{24}
				\Big(  \mc{J}(m_\pi,\D,\mu) - 4\mc{J}(m_K,\D,\mu) + 3\mc{J}(m_\eta,\D,\mu) \Big)
\nonumber\\&\qquad\qquad
				+\frac{3}{2}(m_s - m_l) \Big( \mc{J}(m_\eta,\D,\mu) - \mc{J}(m_K,\D,\mu) \Big)
\nonumber\\&\qquad\qquad
				-\frac{2}{3}(m_s - m_l) \left( m_K^2 \ln \left(\frac{m_K^2}{\mu^2} \right)
					-m_\pi^2 \ln \left( \frac{m_\pi^2}{\mu^2} \right)
				\right)
\nonumber\\&\qquad\qquad
			+\frac{16m_l + 5m_s}{72} 
				\Big(m_\pi^2 -4m_K^2 +3m_\eta^2 \Big)
			-\frac{m_s - m_l}{6} \Big( m_\eta^2 - m_K^2 \Big)
		\bigg]
	\bigg\}
\end{align}
where the function $\mc{J}(m,\D,\mu)$ encodes additional non-analytic dependence on the light quark masses
\begin{multline}
\mc{J}(m,\D,\mu) = 
	2\D\sqrt{\D^2 - m^2+i\e} \ln \left( \frac{\D+\sqrt{\D^2 - m^2 + i\e}}{\D - \sqrt{\D^2 - m^2 + i\e}} \right)
\\
	+m^2 \ln \left( \frac{m^2}{\mu^2} \right)
	-2\D^2 \ln \left( \frac{4\D^2}{m^2} \right)\, .
\end{multline}
and has the limits and properties
\begin{align}
\mc{J}(0,\D,\mu) &= 0
\nonumber\\
\mc{J}(m,0,\mu) &= m^2 \ln \left(\frac{m^2}{\mu^2} \right)
\nonumber\\
\mc{J}(m,-\D,\mu) &= \left\{ \begin{array}{lc}
	\mc{J}(m,\D,\mu) +4i\pi \D \sqrt{\D^2-m^2},& m < |\D| \\
	\mc{J}(m,\D,\mu) -4\pi \D \sqrt{m^2 - \D^2}, & m > |\D|
	\end{array}\right. \, .
\end{align}

%
\section{Evidence for Non-analytic Light Quark Mass Dependence in Baryon Spectrum}
%
%
\subsection{Details of the lattice results}
%
For this work, the numerical results of Ref.~\cite{WalkerLoud:2008bp} are utilized, which are not the most recent but are still the most statistically precise data set available.
The lattice calculation was performed with a mixed-action composed of domain-wall fermion~\cite{Kaplan:1992bt,Shamir:1992im,Shamir:1993zy,Shamir:1998ww,Furman:1994ky} propagators generated on the $n_f = 2+1$ asqtad-improved~\cite{Orginos:1998ue,Orginos:1999cr}, rooted, staggered sea quark configurations generated by the MILC Collaboration~\cite{Bernard:2001av}.
This particular mixed-action set up has been used quite extensively by the LHP~\cite{Renner:2004ck,Edwards:2005kw,Edwards:2005ym,Hagler:2007xi,WalkerLoud:2008bp,Bratt:2010jn} and NPLQCD~\cite{Beane:2005rj,Beane:2006mx,Beane:2006pt,Beane:2006fk,Beane:2006kx,Beane:2006gj,Beane:2006gf,Beane:2007xs,Beane:2007uh,Beane:2007es,Detmold:2008fn,Beane:2008dv,Detmold:2008yn,Detmold:2008bw,Torok:2009dg,Beane:2011zm} Collaboraitons as well as some independent works~\cite{Lin:2007ap,Lin:2008mr,Lin:2009rx,Liu:2009jc,Aubin:2009jh}.
The mixed-action effective field theory, which encodes the discretization effects specific to this particular mixed-action, has also been thoroughly developed~\cite{Bar:2002nr,Bar:2003mh,Bar:2005tu,Golterman:2005xa,Tiburzi:2005is,Chen:2005ab,Prelovsek:2005rf,Aubin:2006hg,Chen:2006wf,Jiang:2007sn,Orginos:2007tw,Chen:2007ug,Aubin:2008wk,Chen:2009su}.  However, the baryon spectrum results we use in this work exist at only a single lattice spacing.  There is also reason to believe the discretization systematics are small~\cite{WalkerLoud:2008bp,Chen:2006wf,Chen:2007ug} and to the order we are working in the mixed-action EFT, they are subleading.  For these reasons, we only use the continuum $\chi$PT extrapolation formula, presented in the previous section.

To set the scale, we use the latest scale setting by the MILC Collaboration~\cite{Bazavov:2009bb}, as detailed in Ref.~\cite{Beane:2011zm}; we first convert the numerical results of Ref.~\cite{WalkerLoud:2008bp} into $r_1$ units%
\footnote{The length scale $r_1$ is determined with the heavy quark potential, defined such that $r_1^2 F(r_1) = -1$.} 
and then use the MILC determination of $r_1(m_l^{phy},m_s^{phy})$ to convert to physical units.
Finally, we perform the extrapolations as functions of the quark masses.  The quark masses are not renormalization scheme or scale independent.  However, at a fixed lattice spacing, we can absorb the quark mass renormalization into the quantity $B$, where at leading order, the Nambu-Goldstone boson masses are given by $m_{i,j}^2 = B(m_i+m_j)$.
We then define lattice quark masses, in physical units by
\begin{equation}
	r_1^{phy} m_q^{latt} \equiv \frac{r_1}{a} (a m_q + a m^{res})\, ,
\end{equation}
where $m^{res}$ is the residual chiral symmetry breaking present with the Domain-Wall lattice action at finite fifth dimensional extent~\cite{Blum:2000kn}.
We collect all these numerical values in Table~\ref{tab:latt-data}.
\begin{table}
\caption{\label{tab:latt-data}\textit{The numerical lattice data from Ref.~\cite{WalkerLoud:2008bp}, converted to physical units.  The uncertainties are the statistical and systematic uncertainties from Ref.~\cite{WalkerLoud:2008bp} combined in quadrature.  The strange quark mass is fixed to the largest $am_q$ value in all calculations: $am_s^{sea} = 0.050$: $am_s^{val} = 0.081$.}}
\begin{ruledtabular}
\begin{tabular}{r|cccccc}
$\beta$& $6.76$& $6.76$& $6.79$& $6.81$& 6.83& 6.85 \\
$am_l^{sea}$& 0.007& 0.010& 0.020& 0.030& 0.040& 0.050  \\
$am_l^{val}$ & 0.0081& 0.0138& 0.0313& 0.0478& 0.0644& 0.081\\
$am_l^{res}$& 0.00160(3)& 0.00157(1)& 0.00123(1)& 0.00101(1)& 0.00083(2)& 0.00073(3) \\
\hline
$m_q^{latt}$ [MeV]& 16.8& 26.6& 58.0& 88.8& 121& 155 \\
$m_\pi$ [MeV]& 320(2)& 389(2)& 557(1)& 685(2)& 805(4)& 905(2) \\
$m_K$ [MeV]& 640(2)& 659(2)& 726(1)& 787(2)& 852(4)& 905(2) \\
\hline
$\frac{3}{2}R_1$ [MeV]& 1285(6)& 1315(6)& 1454(6)& 1556(12)& 1698(13)& 1769(9)\\
$R_3$ [MeV]& -113(3)& -100(2)& -64(1)& -41(1)& -19(1)& 0 \\
$R_4$ [MeV]& -39(2)& -33(1)& -19(1)& -11(1)& -4.4(0.6)& 0\\
$\Delta_\textrm{GMO}$ [MeV]& 5.6(2.3)& 1.8(1.2)& 0.18(48)& 0.13(35)& 0.13(0.09)& 0
\end{tabular}
\end{ruledtabular}
\end{table}

\begin{figure}
\begin{tabular}{cc}
\includegraphics[width=0.48\textwidth]{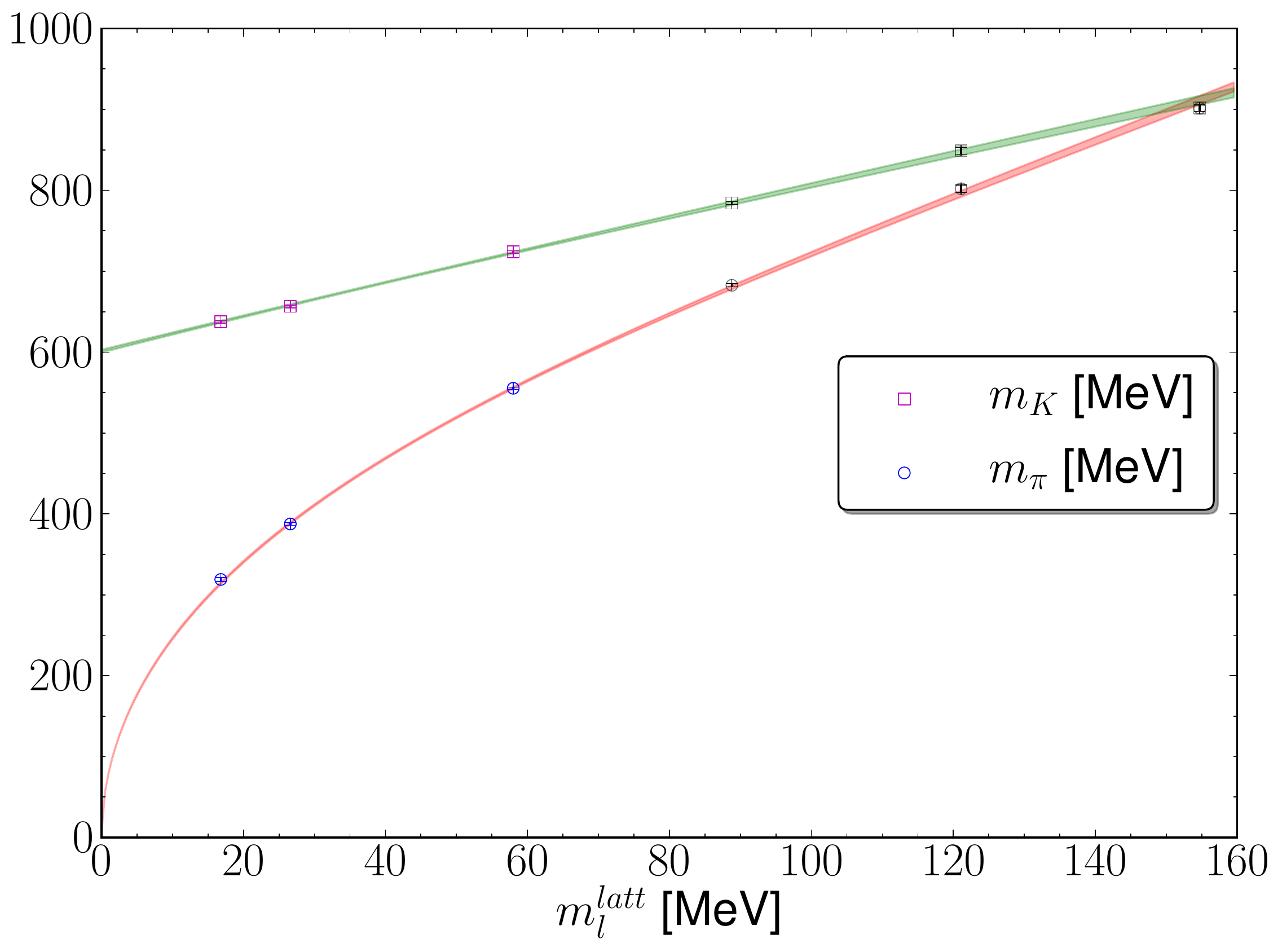}
&
\includegraphics[width=0.48\textwidth]{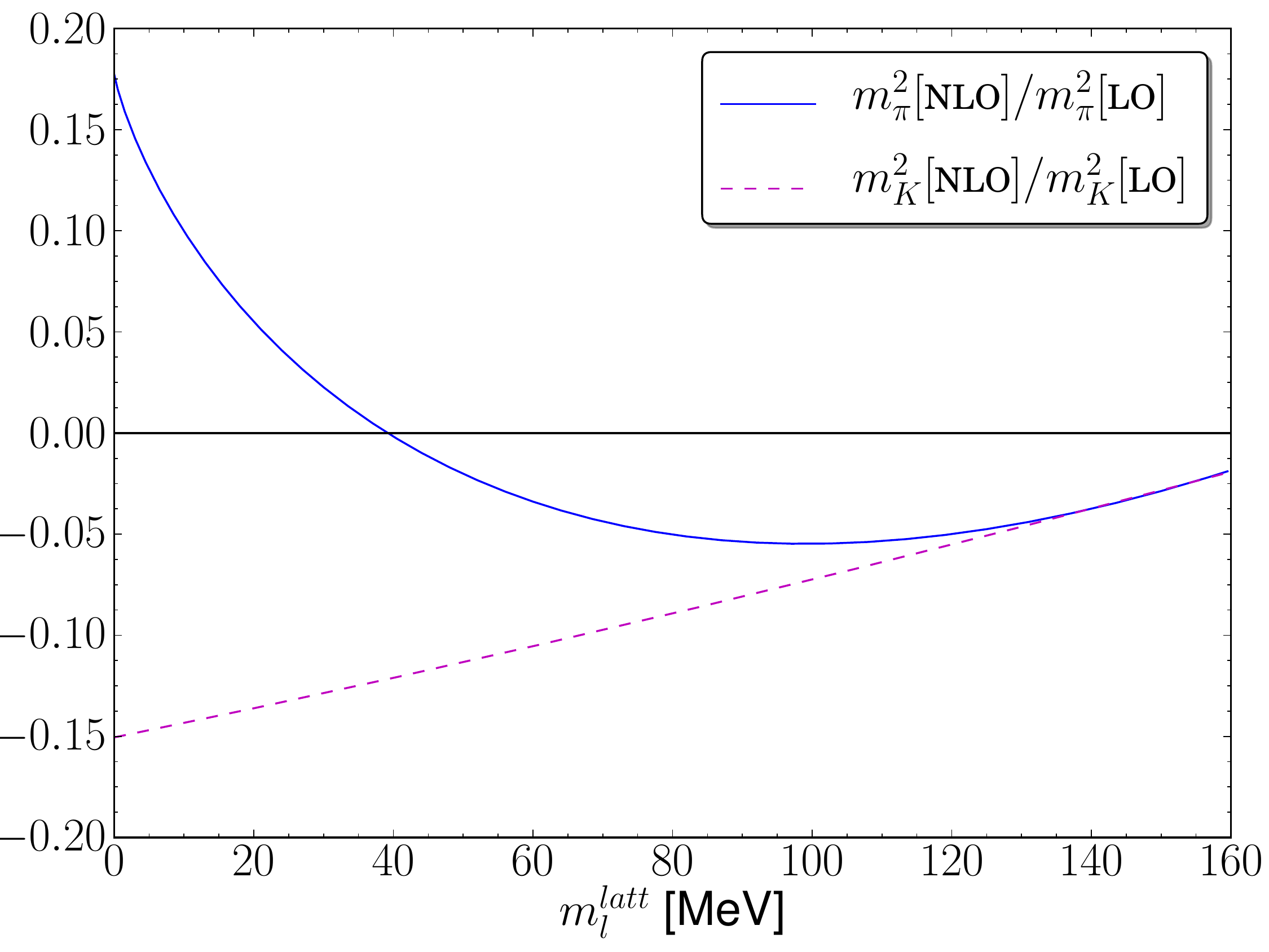}
\end{tabular}
\caption{\label{fig:mpimK}\textit{The results of $SU(3)$ $\chi$PT analysis of the pion and kaon masses.  The left plot displays the fit result with the data and the right plot shows the size of the NLO contributions compared to LO.  This particular fit is performed to just the lightest three values of the light quark mass, and it is clear from the plot that the fits including all six points yields consistent restuls.}}
\end{figure}

To extrapolate the lattice results to the physical point, NLO $\chi$PT~\cite{Gasser:1984gg} is used to determine the values of $m_q^{latt}$ which reproduce
\begin{align}
&m_\pi^{phy} \equiv 138 \textrm{ MeV}\, ,&
&m_K^{phy} \equiv 496 \textrm{ MeV}\, .&
\end{align}
It is interesting to note that despite ignoring the issues of quark mass renormalization, this yields the values
\begin{align}\label{eq:mq_latt_phys}
&m_{l,phy}^{latt} = 3.0(2) \textrm{ MeV}\, ,&
&m_{s,phy}^{latt} = 99(5) \textrm{ MeV}\, ,&
\end{align}
which are remarkably similar to the proper lattice determination of the light and strange quark masses~\cite{Colangelo:2010et}.
The NLO $\chi$PT formula provide a controlled and convergent description of both $m_\pi$ and $m_K$ over the full range of quark masses used, see Fig.~\ref{fig:mpimK}.

%
\subsection{Large $N_c$ and Consistency of Hyperon Axial Charges}
%
One of the major failings in the application of $SU(3)$ heavy baryon $\chi$PT is a lack of consistency between the determination of the axial coupling constants, $D$, $F$, $C$ and $H$ when determined from the baryon spectrum~\cite{WalkerLoud:2008bp,Ishikawa:2009vc} versus a direct calculation of the hyperon axial charges~\cite{Lin:2007ap}.  The direct lattice  determination%
\footnote{In Ref.~\cite{Lin:2007ap}, $g_A$, $g_{\S\S}$ and $g_{\Xi\Xi}$ were computed which were used to infer the values of $D$ and $F$.}
 yields values consistent with the phenomenological values~\cite{FloresMendieta:1998ii}, while the indirect determination from the baryon spectrum yields values consistent with zero.  The small values of the axial couplings returned indicate the numerical results do not support evidence for the leading non-analytic light quark mass dependence predicted in the spectrum.
This problem is not unique to the $SU(3)$ heavy baryon $\chi$PT extrapolations, with large contributions from kaon and eta loops, but also observed in the $SU(2)$ extrapolation of just the nucleon mass.
As demonstrated in Refs.~\cite{WalkerLoud:2008bp,WalkerLoud:2008pj}, for $m_\pi \gtrsim 300$~MeV, there are large cancellations between the LO, NLO and NNLO contributions to the nucleon mass; in order to accommodate the large negative mass contribution occurring at NLO, the leading non-analytic light quark mass dependence, there must be a compensating large but positive contribution from the LO and NNLO terms, signaling a breakdown of the perturbative expansion for these heavier pion masses.  A similar and more severe situation occurs for the $SU(3)$ chiral expansion.
As we shall demonstrate in the next sections, it may be we are simply asking the wrong questions of heavy baryon $\chi$PT.

%
\subsubsection{Mass relation $R_1$}
\begin{figure}
\begin{tabular}{cc}
\includegraphics[width=0.48\textwidth]{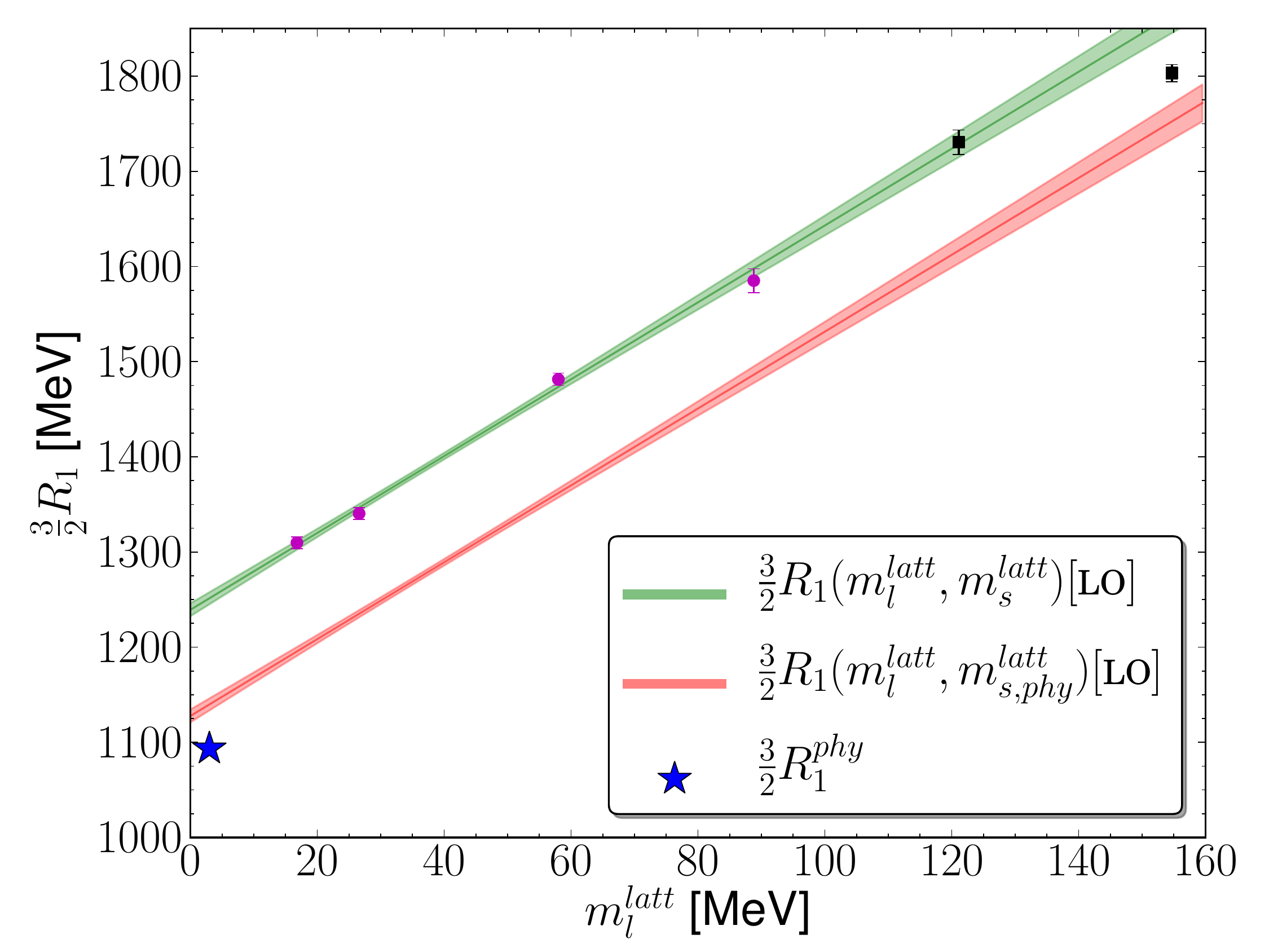}
&\includegraphics[width=0.48\textwidth]{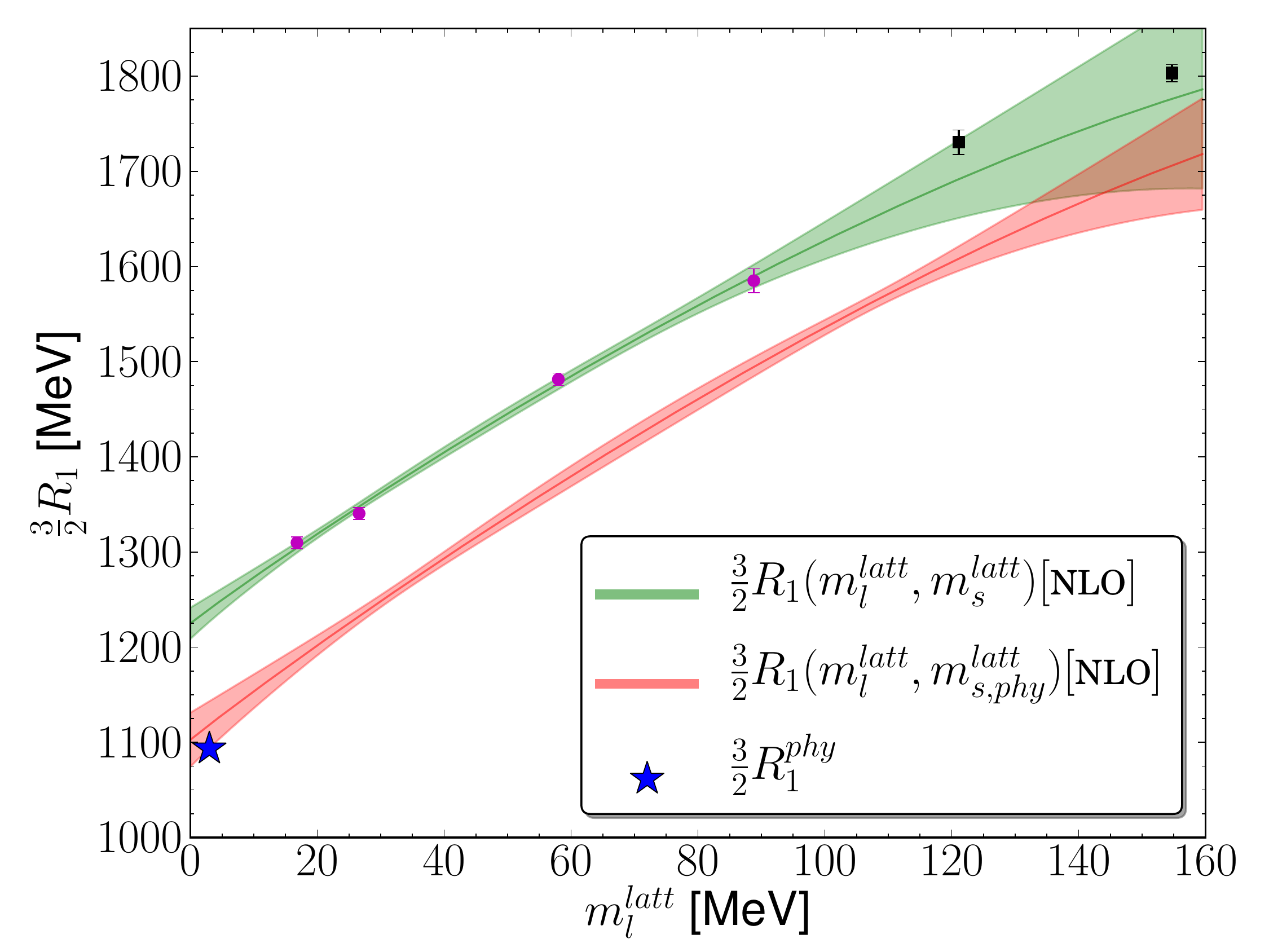}
\end{tabular}
\caption{\label{fig:R1} \textit{Representative fits to $R_1$ from LO (left) and NLO (right) HB$\chi$PT analysis.  The blue star is the physical value, not used in the analysis.  The upper error band results from a fit to the lightest four numerical data and the lower bad is the result extrapolated to the physical value of $m_s^{latt}$. Eq.~\eqref{eq:mq_latt_phys}.}}
\end{figure}

%
The mass relation $R_1$ is a flavor singlet mass combination designed to isolate the $M_0$ contribution plus higher order chiral corrections.
Starting with the lightest quark mass and including successively heavier values of $m_q$, there are four possible ranges of light quark masses which can be used to perform the chiral extrapolation analysis.  
Both LO and NLO analyses are performed over all these ranges of light quark masses.  
In the NLO analysis, the subleading in $N_c$ axial coefficient is set to zero, $a_2 = 0$.
Several choices of the parameter $f$ are taken to explore systematics from higher orders in the chiral expansion:
\begin{align}\label{eq:f_choice}
&i)& &f = f_\pi(m_q)\, ,&
\nonumber\\
&ii)& &f^2 = f_K(m_q)\, f_\pi(m_q) \, ,&
\nonumber\\
&iii)& &f = f_K(m_q)\, ,&
\end{align}
From the LO analysis, the following LECs are obtained
\begin{align}
&M_0[\textrm{LO}] = 903(20) \textrm{ MeV}\, ,&
&\left[b_1 + \frac{5}{18}b_2\right][\textrm{LO}] = -2.73(11)\, .&
\end{align}
Extrapolating to the physical values of the light and strange quark masses gives
\begin{equation}
\frac{3}{2}R_1[\textrm{LO}] = 1124(18) \textrm{ MeV}\, ,
\end{equation}
which is to be compared with $\frac{3}{2}R_1^{phy} = 1093$~MeV.
Performing the NLO analysis, the LECs are determined to be
\begin{align}
&M_0[\textrm{NLO}] = 899(40) \textrm{ MeV}\, ,&
&\left[b_1 + \frac{5}{18}b_2\right][\textrm{NLO}] = -3.26(70)\, ,&
&a_1[\textrm{NLO}] = 0.24(30)\, ,&
\end{align}
with a determination
\begin{equation}
\frac{3}{2}R_1[\textrm{NLO}] = 1127(50) \textrm{ MeV}\, .
\end{equation}
In Fig.~\ref{fig:R1}, representative fits of $R_1$ from LO and NLO are displayed.
One may take comfort in the consistent values of the LECs $M_0$ and $b_1 + \frac{5}{18}b_2$ between the LO and NLO analyses.  However, this is not surprising given the small value of $a_1$ determined in the NLO analysis.  This small value is consistent with no contributions from the NLO terms and inconsistent with the known phenomenological determination of the axial coupling.
This is not surprising given the convergence issues observed in the $SU(2)$ extrapolation of the nucleon mass~\cite{WalkerLoud:2008bp,WalkerLoud:2008pj}.
One is left to conclude that the $SU(3)$ heavy baryon $\chi$PT does not provide a controlled, convergent expansion for the mass combination $R_1$ for the range of quark masses used in this work.

%
\subsubsection{Mass relations $R_3$ and $R_4$\label{sec:R3R4}}
%
The relations $R_3$ and $R_4$ both receive leading contributions from flavor-octet mass operators, vanishing in both the $SU(3)$ vector as well as $SU(3)$ chiral limits.  
From these symmetries, the relations $R_3$ and $R_4$ are more sensitive to the non-analytic light quark mass dependence occurring at NLO in the chiral expansion.
As with the analysis of $R_1$, three choices of the parameter $f$ are taken to estimate higher order effects, Eq.~\eqref{eq:f_choice}.
The LO expressions for $R_3$ and $R_4$, Eqs.~\eqref{eq:R3nlo} and \eqref{eq:R4nlo} with $a_i = 0$, do not describe the numerical results well; it is clear higher order contributions are necessary for extrapolations of this data.
At NLO, the analysis of $R_3$ and $R_4$ becomes correlated.  The full covariance matrix is constructed as described in Ref.~\cite{Jenkins:2009wv}.
The numerical results of Ref.~\cite{WalkerLoud:2008bp} are insufficient to constrain both the leading and subleading axial coefficients, and so the analysis is restricted to the set of LECs
\begin{equation}
	\lambda = (b_1, b_2, a_1)\, ,
\end{equation}
with $a_2 = 0$.
From the NLO analysis, the LECs are determined to be
\begin{align}
	&b_1[\textrm{NLO}] = -6.6(5)\, ,&
	&b_2[\textrm{NLO}] = 4.3(4)\, ,&
	&a_1[\textrm{NLO}] = 1.4(1)\, .&
\end{align}
Using the leading large $N_c$ relations with $a_2 = 0$ in Eq.~\eqref{eq:lgNnloA}, this corresponds to 
\begin{align}
	&D = 0.70(5)\, ,&
	&F = 0.47(3)\, ,&
	&C = -1.4(1)\, ,&
	&H = -2.1(2)\, .&
\end{align}
The significance of this is prominent; the large value of the axial coupling is strong evidence for the presence of the non-analytic light quark mass dependence in these mass relations.  Further, this is the first time an analysis of the baryon spectrum has returned values of the axial couplings consistent with phenomenology.%
\footnote{Finding values of the axial couplings consistent with phenomenology has not just been a challenge for lattice QCD, but also observed in large $N_c$ $\chi$PT analysis of the experimentally measured baryon magnetic moments~\cite{FloresMendieta:2009rq,Ahuatzin:2010ef}.}

\begin{figure}
\begin{tabular}{cc}
\includegraphics[width=0.48\textwidth]{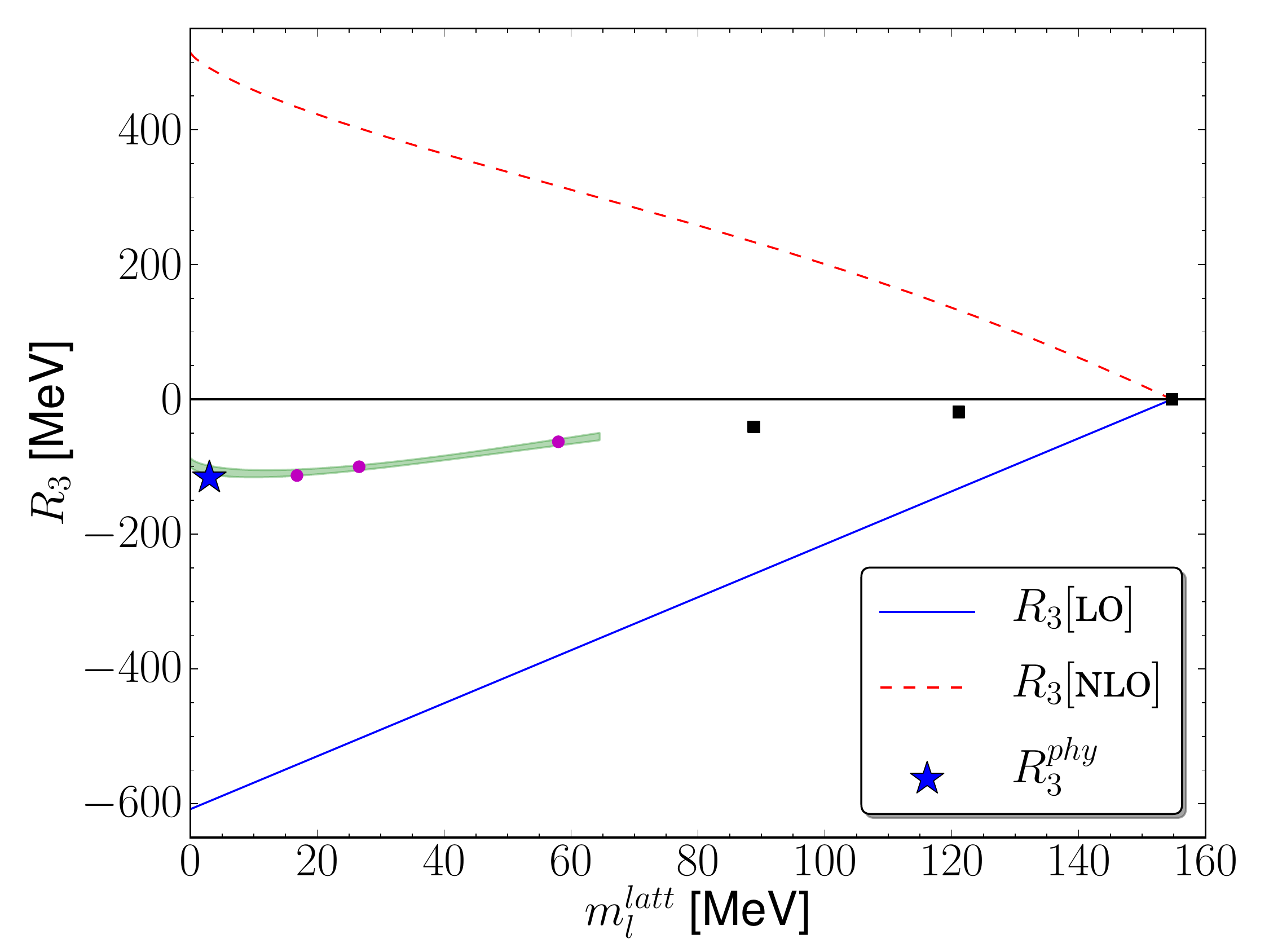}
&\includegraphics[width=0.48\textwidth]{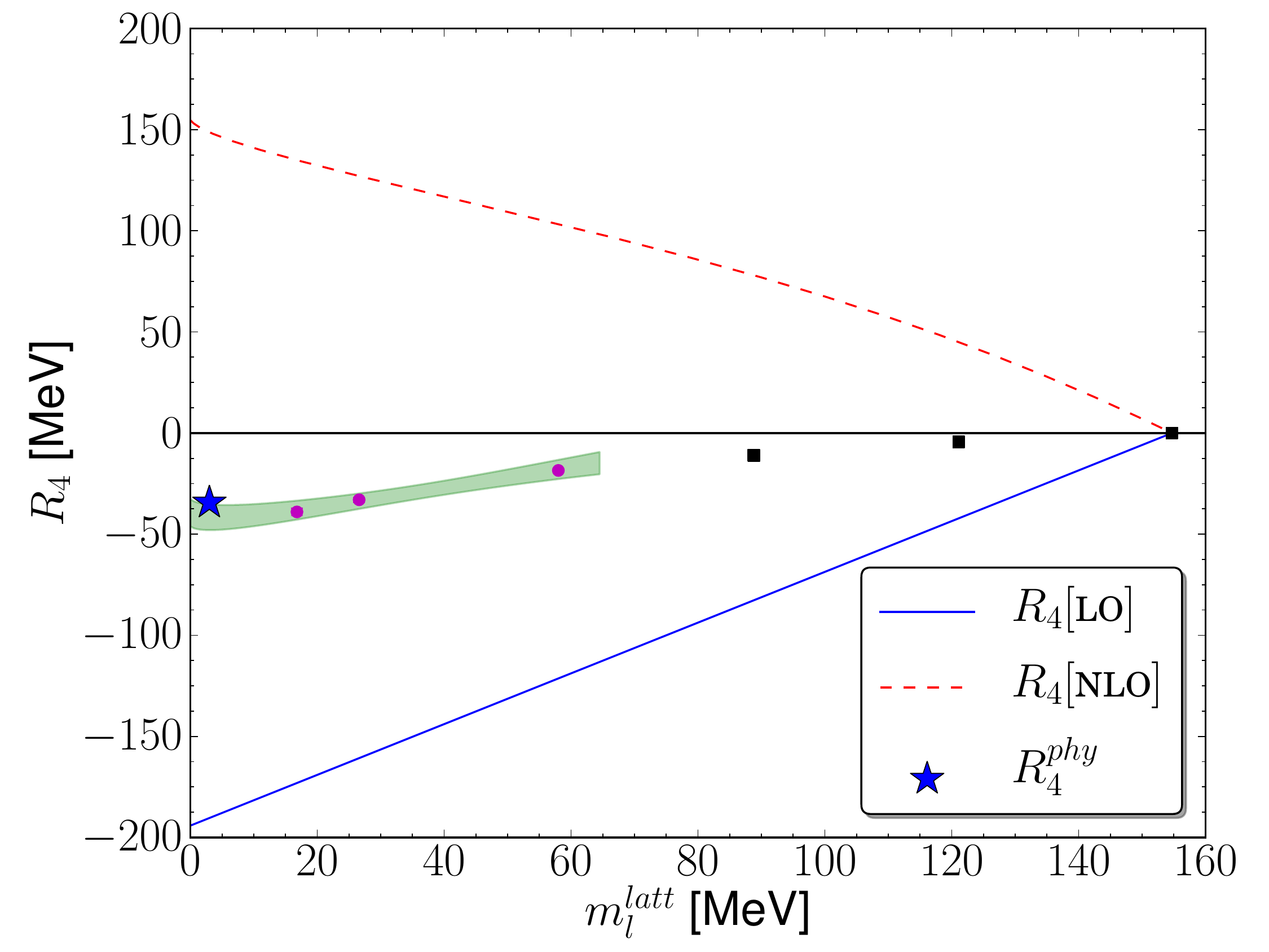}
\end{tabular}
\caption{\label{fig:R3R4_expansion} \textit{The LO and NLO contributions to $R_3$ (left) and $R_4$ (right).  A (blue) star is used to denote the physical values, not included in the analysis.  The particular fit displayed is a combined analysis of $R_3$ and $R_4$ to the data at the lightest three values of $m_l^{latt}$.}}
\end{figure}

However, caution is in order.  Examining the resulting contributions to $R_3$ and $R_4$ from LO and NLO separately, one observes a delicate cancellation between the different contributions, see Fig.~\ref{fig:R3R4_expansion}.  Further studies are needed with more numerical data sufficient to also constrain the parameter $a_2$ as well as the NNLO contributions.

%
\subsection{Gell-Mann--Okubo Relation}
%
The leading contribution to the Gell-Mann--Okubo relation is from a flavor-$\mathbf{27}$, which in HB$\chi$PT come from the leading non-analytic light quark mass dependence, Eq.~\eqref{eq:GMOnlo}.
For this reason, it is a particularly interesting mass relation to study, as has been done if Refs.~\cite{Beane:2006pt,WalkerLoud:2008bp}.
In this article, the analysis is taken a step further.
Close to the $SU(3)$ vector limit, the GMO relation can be described by a taylor expansion in $m_s - m_l$,
\begin{equation}\label{eq:GMOnnloV}
	\D_{\GMO}^V = d_2 \left(m_s - m_l \right)^2 
		+ d_3 \left(m_s - m_l \right)^3
		+ \cdots
\end{equation}
The leading term proportional to $(m_s - m_l)$ must vanish as it transforms as a flavor-$\mathbf{8}$.  The $(m_s - m_l)^2$ contribution is equivalent to an NNLO contribution from HB$\chi$PT and the $(m_s - m_l)^3$ contribution is equivalent to an NNNNLO HB$\chi$PT contribution.
We demonstrate the first few non-vanishing terms in this taylor expansion are inconsistent with the numerical lattice data.
We further demonstrate the NNLO HB$\chi$PT formula can naturally accommodate the strong light quark mass dependence, which is dominated by the non-analytic contributions.
\begin{figure}
\includegraphics[width=0.98\textwidth]{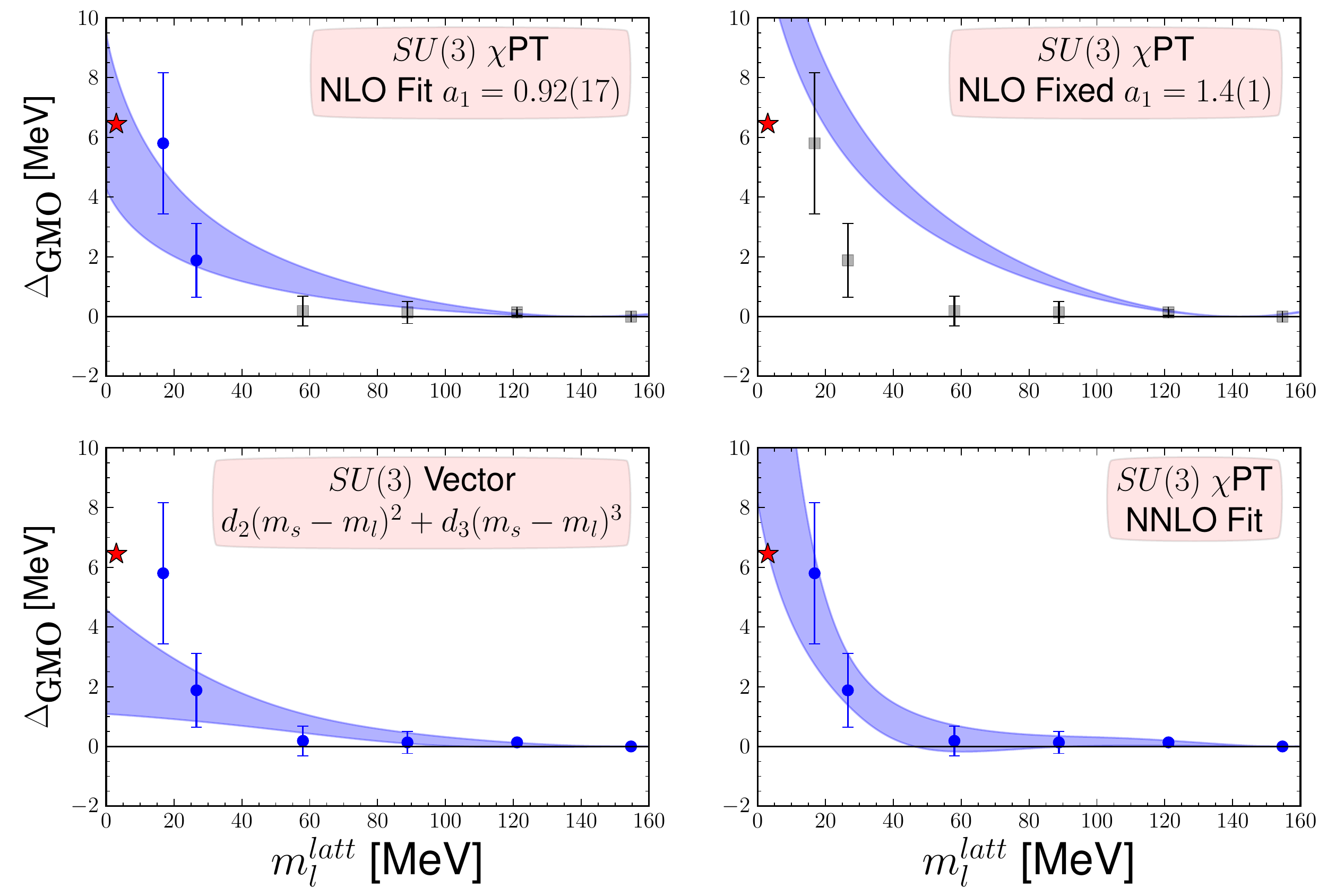}
\caption{\label{fig:GMO} \textit{GMO mass splitting plotted as a function of $m_l^{\rm latt}$.  The $*$ is the PDG point, not included in the analysis.
The various fits are described in the text.
In a given plot, the filled (blue) circles denote results included in the analysis while the open (gray) boxes are excluded.}}
\end{figure}
In Fig.~\ref{fig:GMO}, four plots are displayed.  The first plot (upper left) is the result of an NLO analysis of the GMO formula, allowing the axial coupling to be determined from the data, resulting in a small, but non-zero value for $a_1$.
The second plot (upper right) displays the predicted value of the GMO relation from NLO taking the determination of $a_1$ from the analysis of $R_3$ and $R_4$.
The third plot (bottom left) shows the result of a taylor expansion about the $SU(3)$ vector limit fitting the first two non-vanishing terms.
Finally, the NNLO analysis is displayed, using Eqs.~\eqref{eq:GMOnlo} and \eqref{eq:GMOnnlo} with $a_2 = 0$, and taking $a_1$ from the determination from $R_3$ and $R_4$ (bottom right).
Only the NNLO analysis is consistent with the values of the numerical data over the full range of light quark masses, in particular, the steep rise observed as $m_l^{latt} \rightarrow 0$, as well as the value of the axial coupling $a_1$ determined from phenomenology.
This is further evidence for the presence of non-analytic light quark mass dependence in the baryon spectrum.

%
\section{Conclusions}
%
In this article, we have presented the first substantial evidence for non-analytic light quark mass contributions to the baryon spectrum.
This was achieved by comparing the predictions from heavy baryon $\chi$PT, combined with the large $N_c$ expansion to relatively high statistics lattice computations of the octet and decuplet spectrum.
The numerical results available~\cite{WalkerLoud:2008bp} allowed for a detailed comparison of the mass relations $R_1$, $R_3$ and $R_4$~\cite{Jenkins:1995td} as well as the Gell-Mann--Okubo relation.
It was demonstrated the poor convergence of $SU(3)$ heavy baryon $\chi$PT is isolated in the mass relation $R_1$.
An analysis of mass relations $R_3$ and $R_4$ provided for the first time, values of the axial couplings which are consistent with the phenomenological determination, signaling significant contributions from non-analytic light quark mass dependence in $R_3$ and $R_4$.  At leading order in the large $N_c$ expansion, it was found
\begin{align*}
	&D = 0.70(5)\, ,&
	&F = 0.47(3)\, ,&
	&C = -1.4(1)\, ,&
	&H = -2.1(2)\, .&
\end{align*}
It was further demonstrated that the Gell-Mann--Okubo relation is inconsistent with the first two non-vanishing terms in a taylor expansion about the $SU(3)$ vector limit, and that the steep rise in the numerical data, observed as $m_l^{latt} \rightarrow 0$, can only be described by the NNLO heavy baryon $\chi$PT formula which is dominated by chiral loop contributions.
Taken together, these observations indicate the first significant evidence for the presence of non-analytic light quark mass dependence in the baryon spectrum.

This is not the definitive work however.  There are several known systematics which were not addressed in the present article, and require future, more precise lattice results:
\begin{itemize}
\item the numerical data used~\cite{WalkerLoud:2008bp} exist at only a single lattice spacing,
\item a continuum $\chi$PT analysis was performed,
\item there may be contamination from finite volume effects~\cite{Beane:2011pc},
\item the convergence issues need further examination,
\item more precise numerical results are needed to explore mass relations $R_5$ -- $R_8$ which should be more sensitive to non-analytic light quark mass dependence,
\item results with smaller values of the light quark mass are desireable,
\item the strange quark mass used in this work is known to be $\sim$25\% to large~\cite{Aubin:2004ck}.
\end{itemize}

Addressing these systematics is beyond the scope of this work.  However, current lattice calculations underway should be able to explore these mass relations in more detail.
In particular, the new strategy presented in Ref.~\cite{Bietenholz:2010jr,Bietenholz:2011qq}, where the sum of the quark masses is held fixed, $m_u + m_d + m_s = \bar{m}$, for a range of light and strange quark masses, proves very promising for comparing with predictions from $\chi$PT. 
Further, the strategy is not limited to the spectrum, with similar relations having been recently determined for the baryon magnetic moments~\cite{Jenkins:2011dr}.

\acknowledgments

\noindent
We thank the LHP Collaboration for use of their numerical data~\cite{WalkerLoud:2008bp}.  We thank E.~E.~Jenkins for involvement at early stages of this work and we warmly acknowledge the hospitality of the UCSD High Energy Theory Group where part of this work was completed.
We thank C.~Bernard for useful discussions and the values of $r_1/a$ used for scale setting.
The work of AWL was supported in part by the
Director, Office of Energy Research, Office of High Energy and Nuclear
Physics, Divisions of Nuclear Physics, of the U.S. DOE under Contract
No.~DE-AC02-05CH11231.

\bibliography{largeNcSU3}

\vfill\break\eject

\end{document}